\documentclass[aps,prd,preprint,nofootinbib]{revtex4}
\usepackage{amsfonts}
\usepackage{mathrsfs}
\usepackage{graphicx}
\usepackage{amsmath}
\usepackage{amssymb}
\usepackage{epsfig}
\usepackage{graphicx}
\usepackage{slashed}
\usepackage{color}
\usepackage{multirow}
\usepackage{ulem}
\usepackage{adjustbox}

\usepackage{stmaryrd}
\newcommand{\bqa}{\begin{eqnarray}}
	\newcommand{\eqa}{\end{eqnarray}}
\newcommand{\beq}{\begin{equation}}
	\newcommand{\eeq}{\end{equation}}
\allowdisplaybreaks[1]

\begin{document}

\title{Center of mass motion in  bag model}
\author{ Chia-Wei Liu\footnote[1]{chiaweiliu@ucas.ac.cn} and  Chao-Qiang Geng \\}
\affiliation{School of Fundamental Physics and Mathematical Sciences, Hangzhou Institute for Advanced Study, UCAS, Hangzhou 310024, China\\
	University of Chinese Academy of Sciences, 100190 Beijing, China
	\vspace{0.6cm}} 

\baselineskip=20pt

\begin{abstract}
Despite the great success on the mass spectra, the reputation of the bag model has been closely followed by the embarrassment from the center of mass motion. It leads to severe theoretical inconsistencies. For instance, the masses and the decay constants would no longer be independent of the momentum. In this work, we provide a systematical approach to resolve this problem. The meson decay constants as well as the baryon transition form factors can be computed consistently in our framework. Notably, the form factors in the neutron $\beta$ decays do not depend on any free parameters, and are determined to be $F^V_1 =1 $ and $F^A_1 = 1.31$ or $F_1^A/F_1^V= 1.31$, which is close the experimental value of $F^A_1/F^V_1 = 1.27$. In addition, we find that ${\cal B} (\Lambda_b \to \Lambda \gamma) = (6.8 \pm 3.3 ) \times 10^{-6} $, which agrees to the experimental value of $(7.1\pm 1.7)\times 10^{-6}$. 

\end{abstract}

\maketitle

\section{Introduction}
The Massachusetts Institute of Technology (MIT) bag model
 describes a hadron as a bag which contains various quarks, antiquarks and perhaps  gluons~\cite{Chodos:1974je}.
It was proposed to reconcile two very different ideas, behind the quantum chromodynamics~(QCD) with a single parameter of  the bag radius~($R$),  as follows:
\begin{itemize}
\item Due to the asymptotic freedom,   quarks move freely inside the bag;
\item  Quarks are not allowed to penetrate the bag because of  the QCD confinement. 
\end{itemize}
Expressing the ideas mathematically,
 we start with  the free Dirac equations, given by 
 \begin{eqnarray}\label{free dirac}
(i \slashed{\partial } -M_q)\psi_q (x) = 0\,,~~~~~\text{ in }~r<R \,, 
 \end{eqnarray}
along with the  boundary condition,
 \begin{eqnarray}\label{boundary}
-i \left( \hat {x} \cdot \vec{\gamma } \right) \psi_q(x)    = \psi_q(x) \,,~~~~~\text{ at }~r = R \,,
\end{eqnarray}
where 
$\psi_q$ stands for the quark wave function, 
$M_q$ represents the quark mass with $q$  being  the quark  flavor, $r=|\vec{x}|$,  $\vec{\gamma} =(  \gamma_{1} , \gamma_2, \gamma_3 ) $, and $R$ is   the bag radius. 
The description  is  similar to the infinite square well in quantum physics. 
In practice,  “bag'' can be understood as the abbreviation for “infinite spherical well''. 
The higher order mass corrections can be done by taking $\psi_q$ as an unperturbed state.

Thanks to its simplicity, the bag model can easily cooperate with various QCD systems.
The applications range from atoms to burning stars. In  nuclear physics, the bag model offers an intuitive understandable ground for studying the mass spectra~\cite{cloudy BAG,DeGrand:1975cf,cloudy BAG2,Ref2} and  the   hypothetical objects,  such as   $\Xi_{bb}$, glue balls and pentaquarks
~\cite{Hypothetic,Hypothetic1,Hypothetic2,Hypothetic3}.
 See Ref.~\cite{Review} for a historical review.
In  astrophysics, under the framework of the bag model,  the strange stars, neutron stars and quark cores are  viewed  as large bags containing countless quarks~\cite{MIT Neutron,MIT Neutron1,MIT Neutron2,MIT Neutron3,MIT Neutron4}.
In spite of these great progresses, the bag model has little application in the particle decay system, mainly due to the center of mass motion~(CMM)~\cite{Duck:1976fp,Duck:1976fp1,Duck:1976fp2}.
This study is devoted to introduce  the homogeneous bag model to  provide a consistent framework to deal with the CMM. 

The paper is 
organized as  follows. We give a brief review of the  MIT bag model in Sec.~\MakeUppercase{\romannumeral 2}, where
we concentrate on the problem of the CMM rather than computational details.
In Sec.~\MakeUppercase{\romannumeral 3}, we display some of the methods tackling with the CMM given in the  literature.
In Sec.~\MakeUppercase{\romannumeral 4}, we present  a systematic framework to deal with the CMM, and show   the numerical results of the neutron $\beta$ decay and $\Lambda_b \to \Lambda \gamma$. 
We conclude the study 
in Sec.~\MakeUppercase{\romannumeral 5}.

\section{The MIT bag model}
Since we only consider the low-lying hadrons, 
the quark wave function of 
$\psi_q$ can be safely taken to be isotropic. 
Consequently,  Eq.~\eqref{free dirac}  with $J_z=1/2$ can be easily solved as  
\begin{equation}\label{quark_wave_function}
	\psi_{q\uparrow} (x) =\phi_{q\uparrow}(\vec{x}) e^{-iE_qt}= N\left(
	\begin{array}{c}
		\omega_{q+} j_0(p_qr) \chi_\uparrow\\
		i\omega_{q-} j_1(p_qr) \hat{r} \cdot \vec{\sigma} \chi_\uparrow\\
	\end{array}
	\right)e^{-iE_qt}\,,
\end{equation}
where  $\phi_q(\vec{x})$ corresponds to the spatial part of the quark wave function with $E_q^k = \sqrt{p_q^2  +M_q^2}$\,, $\omega_{q\pm} = \sqrt{E^k_q \pm M_q}$ \,, 
$p_q$ and $E_q$   are the magnitudes of the quark 3-momentum and energy, respectively, $N$ is the normalization constant,   $\chi_\uparrow$ and $\chi_\downarrow$ stand for $(1,0)^T$ and $(0,1)^T$, respectively,  and 
 $j_{0,1}$ are the zeroth and the first spherical Bessel functions, which are basically the cosine and sine functions  in the spherical coordinate. 
For  the absence of energy corrections, we  have  that $E_q = E_q^k$.

In analogy to the case of the infinite square well, imposing the boundary condition would quantize $p_q$. 
Combining Eqs.~\eqref{boundary} and \eqref{quark_wave_function}, we find that $p_q$ must satisfy the relation~\cite{DeGrand:1975cf}
\begin{equation}
\tan \left(p_q R\right)=\frac{p_q R}{1-m_q R-E_q^k R}\,.
\end{equation}
At the massless  and   heavy quark limits, we obtain 
\begin{equation}
\lim_{M_q R \to 0 } p_q R = 2.043, 5.396\cdots \,,~~~ \lim_{M_q R \to \infty } p_q R =   \pi\,,2\pi \cdots  \,,
\end{equation} 
respectively, which is valid for all hadrons. These results  are handy for  quick estimations as $M_{u,d}$  can be taken as massless,  whereas $M_{c,b}$ as infinite in practice. 
It states that the magnitude of the 3-momentum grows along with the quark mass, approaching $ \pi /R$.

At this stage, one can readily give the estimations of the hadron masses by adding up the quark energies. 
We take the proton as an  example,  and the formalism can be easily generalized to other hadrons.
The  proton  wave function is  given by superposing the quark states
\begin{equation}\label{naive}
			\Psi(x_1,x_2,x_3) = \langle 0 | \hat{u}(x_1)\hat{ u}(x_2)\hat{d}(x_3) | p_B(0) \rangle = \psi_u(x_1) \psi_u(x_2) \psi_d(x_3) \,,
\end{equation}
where $\hat{ u}$ and $\hat{d}$ are the quark field operators, without writing down  the spin-flavor and color indices explicitly, and  $|p_B(0)\rangle$ represents the proton state from the bag model, centered in the coordinate. 
The experimental  proton charge radius of $0.840$~fm~\cite{Lin:2021xrc} corresponds to   $R= 5.85$~GeV$^{-1}$ in the bag model,
 leading to the proton and  $\Delta$  masses as 
\begin{equation}
M_{p, \Delta} = \frac{2.043}{5.85 } \times 3 = 1.047~\text{GeV}\,,
\end{equation}
which are  close to the experimental value,
\begin{equation}
\frac{	1 }{2}
\left(
M_p + M_\Delta 
\right)_{\text{EXP}} = 1.085~\text{GeV}\,. 
\end{equation}
However, the neutron charge radius is essentially zero in the bag model,  which is inconsistent with the experiments.

There are several  mass corrections, which can be summarized as follows~\cite{DeGrand:1975cf}:
\begin{itemize}
\item  The energy of the bag is proportional to its volume,  given by  $E_V =  4\pi B R^3/3$ with $B$ the bag energy density.
\item The zero-point energy is found to be negative empirically, given by $E_0 = -Z_0/R$. 
\item  The quark-gluon interaction introduces the strong coupling constant of  $\alpha_s$.
\end{itemize}
Here, $B$, $Z_0$ and $\alpha_s$  are treated as free parameters in the model. On the other hand, the bag radii are determined by minimizing the hadron masses, given as~\cite{DeGrand:1975cf}
\begin{equation}
\left. \frac{\partial M_H}{\partial R}\right|_{R=R_H} = 0\,,
\end{equation} 
where $M_H$ and $R_H$ are the mass and bag radius of $H$, respectively,
resulting in different hadron bag radii. Typically, the baryon  and meson radii are found to be around  $5$ and $3$ GeV${^{-1}}$, respectively.
Except for $u$ and $d$,
the quark masses
are taken to be free parameters in the model, fitted to be~\cite{Hypothetic3}
\begin{equation}\label{extractMass}
	M_s = 0.279~\text{GeV}\,,~~
	M_c =1.641~\text{GeV}\,,~~M_b=5.093~\text{GeV}\,.
\end{equation}
which can be viewed as  the effective masses when the quarks swim inside the bags.

Although the mass corrections from the bag and the zero-point energies are intuitively satisfactory, they are problematic under the spacetime symmetry. 
 For instance,  particles  at rest are invariant under the space translations. How can their masses related to a finite volume? How do the zero-point energies transform under the Lorentz boost? 
These problems partly come from that the description of the bag model is a semi-classical one. 
Quarks are quantum objects as they satisfy the Dirac equation, whereas   the bag itself, having a definite position as well as a concrete boundary, is a classical object.
These are the essential problems that have haunted the bag model ever since it was proposed, closely related to the CMM. 
As a pioneer of quark models, it provides an excellent framework to understand the hadrons in relativistic systems. However, it can hardly be applied to the particle decay processes.

Another  issue after considering the mass corrections is that we lost the ability to track the $t$-dependencies of the individual quarks. The best we can get is the $t$-dependency of the proton wave  function, read as 
\begin{equation}\label{second naive}
	\Psi(t, \vec{x}_{1},\vec{x}_{2},\vec{x}_{3})  =e^{-iM_p t } \phi_u(\vec{x}_1)\phi_u(\vec{x}_2)\phi_d(\vec{x}_3)\,.
\end{equation}
Here, we can no longer take $t_{1,2,3}$ to be  different as done in  Eq.~\eqref{naive} for $2E^k_u + E^k_d\neq M_p$ after including the bag and zero-point energies.
Due to this reason, we can not apply a Lorentz boost on Eq.~\eqref{second naive}\,, which would mix  up the spacetime coordinates. The problem can be traced back to that the $t=0$ plane is not invariant under a Lorentz boost, which we will discuss more details in Sec.~\MakeUppercase{\romannumeral 4}, in which the wave functions are spell out in terms of the creation operators.

\section{The wave packet and ellipsoidal bag approaches}
We are interested in the applications of the bag model in the decay system, in  which momentum eigenstates are required. 
In the literature, there are 
two methods concerning decays with the bag model.
These methods have their own advantages but fail to achieve a consistent framework
with the Poincar\'e symmetry. Nonetheless, they  shed light on the CMM problem and  we discuss them briefly.
\subsection*{A bag as a wave packet}
The most na\"ive solution to explain the CMM is 
the wave packet approach, which
treats the   bag wave function as  a localized wave packet. We {\it extract} the momentum states based on the  Fourier analysis, read as 
\begin{equation}\label{wavepacket}
|p ( \vec{p})\rangle =N(\vec{p})  \int d^3x ~ e^{- i \vec{p}\cdot \vec{x}} |p_B (\vec{x})\rangle\,, 
\end{equation}
where the left hand side is a proton state  with $\vec{p}$ the 3-momentum, leading to the quark state decomposition as 
\begin{align}
	\Psi(t, \vec{x}_1,\vec{x}_2,\vec{x}_3) 
=N(\vec{p}) \int d^3x e^{-i \vec{p}\cdot \vec{x} -iM_p t } \phi_u(\vec{x}_1- \vec{x})\phi_u(\vec{x}_2- \vec{x} )\phi_d(\vec{x}_3- \vec{x })\,.
\end{align}
One of the advantages  of this method is that the $t$-dependencies of the quark states are not required, so it can easily cooperate with the bag and zero-point energies\footnote{See the discussion following Eq~\eqref{second naive}.}.

\subsection*{The ellipsoidal bag approach}
The contradiction occurred in the wave packet approach is attributed to that  we can not have a wave packet with a definite energy. To solve the problem, 
the
 Lorentz boost  for  the bag state is needed~\cite{Barnhill:1978ui,Barnhill:1978ui1,Barnhill:1978ui2,Barnhill:1978ui3}. 
Boosting Eq.~\eqref{naive},    the  wave function is read as 
\begin{equation}\label{13}
			\Psi ^v (x_1,x_2,x_3)  = \psi^v_{u}(x_1)\psi^v_{u}(x_2)\psi^v_{d}(x_3)
	\end{equation}
with $v$ the velocity and 
\begin{equation}\label{dem}
 \psi_q^v(x) =  S_v \psi_q(\Lambda_v^{-1}  x) \,,
\end{equation}
where  $S_v$ and 
$\Lambda_v$ are the Lorentz boost matrices toward $\hat{z}$ direction for Dirac spinors and  spacetime coordinates, respectively.
In this work $\vec{v}$ is always chosen in the $\hat{z}$ direction, leading to $S_v = a_+ + a_-\gamma^0\gamma^3$ with $a_\pm=\sqrt{\gamma\pm1}$ and $\gamma ^{-1}= \sqrt{1-v^2}$.
In contrast to Eq.~\eqref{naive},
the spherical bag deforms to an ellipsoid due to the Lorentz contraction, and the bag itself is moving. 

Clearly,    the $t$-dependencies~($E_q$) of the quark states are needed by Eq.~\eqref{dem}.
A reasonable range for the up  and down quark energies is given as 
\begin{equation}\label{energy}
\qquad\qquad\frac{1}{3} M_p<E_{u,d} <  E_{u,d}^k + \frac{1}{3} \left(
E_0 + E_V
\right)\,,
\end{equation}
or numerically
\begin{equation}\label{energyn}
0.313~\text{GeV}<E_{u,d} < 0.368~\text{GeV}\,,\qquad
\end{equation}
where we  allocate $E_0$ and $E_V$ evenly among the quarks. Notice that 
 Eq.~\eqref{13} admits implicitly that the energies of the quarks are independent to each others, which is not true if there have interactions among them.
 As a consequence, we have  $E_{u,d}  >M_p/3$, and Eq.~\eqref{energyn} serves as a major uncertainty in the evaluation. Nonetheless, for the processes where the initial and final baryons have the same velocity, the Lorentz boost is not required and therefore the calculation does not suffer from the uncertainty.

If we take Eq.~\eqref{13} as a momentum eigenstate, there are couple requirements  to be added  by hand before the calculations : 
\begin{itemize}
\item To have the same bag volume, the initial and  final hadrons must be  opposite in velocities.
\item As the bag is moving,  a specific timing  for the computation has to be chosen. 
\end{itemize}
The studies of the ellipsoidal bag approach are  carried out in  the decays with the $b \to c $ transition~\cite{Sadzikowski:1993iv}, where the overlapping between the heavy quarks is treated by  the heavy quark symmetry. 
On the other hand,
the overlapping of the light quarks would lead to the $v$-dependency of the Isgur-Wise function, read as 
\begin{equation}\label{Dv10}
\langle l_v |  l_{-v}\rangle  = \int d^3\vec{x} \psi^{v\dagger }_l(0, \vec{x})\psi^{- v }_l(0, \vec{x}) =  D^v_l (0)  \,,
\end{equation}
with 
\begin{equation}\label{Dvdefin}
D^v_l (\vec{x}_\Delta ) = \frac{1}{\gamma}
\int d^3 \vec{x } \phi_l^\dagger\left(\vec{x}+ \frac{1}{2} \vec{x}_\Delta \right)  \phi_l \left(\vec{x}- \frac{1}{2} \vec{x}_\Delta\right)  e ^ { - 2 i E_l\vec{v}\cdot \vec{x}} \,,
\end{equation}
where
 $l$ is the spectator quark in the heavy  hadron transition, and $D^v_l(\vec{x}_\Delta )$ is a function of both velocity and position, where the dependence on $\vec{ x}_\Delta$ is defined for the latter convenience. 
In Ref.~\cite{Sadzikowski:1993iv}, it was  found that ${\cal B}( B^0  \to D^-  l^+{\nu}_l) = 2.12\%$, which is close to  $(2.31\pm 0.10)\%$ given by the experiments~\cite{pdg}. 
Remarkably,  Eq.~\eqref{Dv10}  can be understood intuitively, in which $\gamma^{-1}$ comes from the Lorentz contraction of the bag volume, while the exponential in the integral  causes the  damping, which is a punishment for not being at the same velocity.

However,
it is unarguable that both the Lorentz and translational symmetries are broken in the ellipsoidal bag approach, since a specific initial frame and timing are demanded in the computation.
Furthermore, Eq~\eqref{Dv10} tell us that the inner products between $|l_{\pm v}\rangle$  are nonzero,  violating  the energy momentum conservation.

\section{ The homogeneous bag model }

The homogeneous bag model, first proposed  in Ref.~\cite{Geng:2020ofy},  has been widely  applied in various decay systems~\cite{various,various1,various2,various3}.
It is meant to reconcile the Poincar\'e symmetry and the bag model, without 
necessarily
introducing  a new parameter.
However, in Ref.~\cite{Geng:2020ofy} only the scalar operators were considered. In this work, we would like to generalize the formalism to the vector and tensor operators as well.

In the following discussion, we will  construct hadron states by combining the features of the wave packet and  ellipsoidal bag approaches.
Remind that the wave packet approach violates the Lorentz symmetry as masses depend on  velocities, whereas the ellipsodial bag one breaks the translational symmetry as the wave functions are localized. 
We will show that the inconsistencies with the Poincar\'e symmetry are resolved in our  framework. We also take the neutron $\beta$ decay for an instance of the computation, as it is independent of free parameters.

\subsection*{Hadron wave functions}
We start with a hadron state at rest. As a zero momentum state has to  distribute homogeneously over the space, we linearly superpose infinite bags at different locations. A proton state at rest is {\it constructed} as 
\begin{equation}
|p (\vec{p} = 0)  \rangle = {\cal N}_p  \int d^3 \vec{x} | p_B(\vec{x})\rangle \,,
\end{equation}
where ${\cal N}_H$ is the normalization constant for the hadron $H$. 
Notice that the formula is identical to Eq.~\eqref{wavepacket} when  $\vec{p}=0$, but the idea is totally different. 
Eq.~\eqref{wavepacket} is meant to extract the specific momentum component from the wave packet, whereas we are trying to build up a momentum eigenstate from infinite static bags here.

For completeness, we would  write down the baryon wave functions with color and spinor indices. It can be accomplished by adopting the quark creation operators, given by 
\begin{equation}\label{19}
|p (\vec{p}= 0 ),\uparrow \rangle =  \int [d^3\vec{x}] 
\frac{1}{2\sqrt{3}}
\epsilon^{\alpha \beta \gamma} u^\dagger_{a\alpha}(\vec{x}_1)d^\dagger_{b\beta}(\vec{x}_2)u^\dagger_{c\gamma}(\vec{x}_3)	\Psi_{A_\uparrow (udu)}^{abc} ( \vec{x}_1, \vec{x}_2 ,  \vec{x}_3 )|0\rangle \,,
\end{equation}
along with
\begin{equation} \label{wave_function_inrest}
	\begin{aligned}
		\Psi_{A_{\updownarrow}(q_1q_2q_3)}^{abc} ( \vec{x}_1, \vec{x}_2 ,  \vec{x}_3 ) =& \frac{{\cal N}_H}{\sqrt{2}} \int \left [ \phi^a_{q_1\uparrow}(\vec{x}_1- \vec{x}_\Delta) \phi^b_{q_2\downarrow}(\vec{x}_2-\vec{ x}_\Delta) \right. \\
		&\left. - \phi^a_{q_1\downarrow}(\vec{x}_1- \vec{x}_\Delta) \phi^b_{q_2\uparrow}(\vec{x}_2-\vec{ x}_\Delta) \right]  \phi^c_{q_3\updownarrow}(\vec{x}_3- \vec{x}_\Delta) d^3 \vec{x}_\Delta,
	\end{aligned}
\end{equation}
where $[d^3\vec{x}]= d^3\vec{x}_1d^3\vec{x}_2d^3\vec{x}_3$, 
the Greek~(Latin) letters stand for the color~(spinor) indices,  
 the arrows represent  the spin directions of the quarks, and the Fermi statistic is guaranteed by the anti-commutation relation
 \begin{equation}\label{21}
\{ q_{a\alpha}(\vec{x})  , q^{\prime \dagger }_{b\beta}(\vec{x}\,')  \} =\delta_{qq'} \delta_{ab}\delta_{\alpha\beta} \delta^3\left(
\vec{x} - \vec{x}\,'
\right)\,.
 \end{equation}
For the sake of compactness, the quarks operators are evaluated at $t=0$ if not stated otherwise.

To obtain a nonzero momentum, we  have to apply the Lorentz boost $U_v$ on Eq.~\eqref{19}.  
Recall that
the transformation rule of the quark operators is  given as 
\begin{equation}\label{Lorentz}
U_v^{-1} q _{a \alpha } (  \vec{x}  ) U_v  = \left( S_v  \right) _{ab} q _{b \alpha } \left(  x^v\right )\equiv  \left( S_v  \right) _{ab} q _{ b \alpha } \left(  t=-\gamma v z , x, y , \gamma z \right ) \,.
\end{equation}
It states that even if we start with the operators at  an equal time, we inevitably have to deal with the quark operators with timelike distances  as the quarks have different positions in the baryon at rest. The problem can be traced back to that the plane $t=0$ is not invariant under the Lorentz boost, which leads to a unequal time among the operators. 
The unequal time commutators require the knowledge of the dynamical details, which can not be perturbatively calculated. To overcome the problem, we utilize that
\begin{equation}\label{23}
q_{a\alpha }( t, \vec{x }_i) | p (\vec{p} = 0 ) \rangle = q_{a\alpha }( 0, \vec{x }) e^{-iE_q t} | p (\vec{p} = 0 ) \rangle \,,
\end{equation}
which stem from that the quarks are energy eigenstates  in the bag model at least for the first order approximation. 

The wave functions after the Lorentz boost can be obtained by the following trick. Without lost of generality, we write the wave functions after boosting    as 
\begin{equation}\label{24}
U_v|p (\vec{p}= 0 ),\uparrow \rangle =  \int [d^3\vec{x}]\frac{1}{2\sqrt{3} } \epsilon^{\alpha \beta \gamma} u^\dagger_{a\alpha}(\vec{x}_1)d^\dagger_{b\beta}(\vec{x}_2)u^\dagger_{c\gamma}(\vec{x}_3)	( \Psi^v )_{A_\uparrow (udu)}^{abc} ( \vec{x}_1, \vec{x}_2 ,  \vec{x}_3 )|0\rangle \,,
\end{equation}
where $\Psi^v$ is a function to be determined.
Note that as $J_z$ commutes with $U_v$, the proton remains as an eigenstate of $J_z$. 
 Applying the annihilation operators, we arrive at 
\begin{eqnarray}\label{25}
&&\left\langle 0 \left| u_{a\alpha} (\vec{x}_1)d_{b\beta} (\vec{x}_2)u_{c\gamma} (\vec{x}_3) U_v \right| p (\vec{p}= 0 )\right\rangle \nonumber\\
&=&\left\langle 0 \left| U_v  U_v ^{-1} u_{a\alpha} (\vec{x}_1)d_{b\beta} (\vec{x}_2)u_{c\gamma} (\vec{x}_3) U_v \right| p (\vec{p}= 0 )\right\rangle\nonumber\\
&=&\left\langle 0 \left| U_v ^{-1} u_{a\alpha} (\vec{x}_1)U_v  U_v ^{-1}d_{b\beta} (\vec{x}_2)U_v  U_v ^{-1}u_{c\gamma} (\vec{x}_3) U_v \right| p (\vec{p}= 0 )\right\rangle\\
&=&\left\langle 0 \left| (S_v)_{aa'} u_{a'\alpha} \left( x^v_1\right )(S_v)_{bb'} d_{b'\beta} \left( x^v_2 \right )(S_v)_{cc'} u_{c'\gamma} \left( x_3^v \right )  \right| p (\vec{p}= 0 )\right\rangle\nonumber\\
&=&e^{i\gamma v (E_uz_1+E_d z_2 + E_u z_3)}
(S_v)_{aa'}(S_v)_{bb'}(S_v)_{cc'}
\left\langle 0 \left|  u_{a'\alpha} \left( \vec{x}^v_1\right ) d_{b'\beta} \left( \vec{x}^v_2 \right ) u_{c'\gamma} \left( \vec{x}_3^v \right )  \right| p (\vec{p}= 0 )\right\rangle\,,\nonumber
\end{eqnarray}
with $\vec{x}^v = (x,y,\gamma z)$. In the third line of Eq.~\eqref{25}, we have used that the vacuum is invariant under Lorentz boosts, and the fourth and fifth ones can be obtained by Eqs.~\eqref{Lorentz} and \eqref{23}, respectively.
By comparing the first and fifth lines of Eq.~\eqref{25}, we deduce that 
\begin{eqnarray}\label{26}
(\Psi^v)_{A_\uparrow(udu)} ^{abc} (\vec{ x}_1,\vec{ x}_2,\vec{ x}_3)=e^{i\gamma v (E_uz_1+E_d z_2 + E_u z_3)} (S_v)_{aa'}(S_v)_{bb'}(S_v)_{cc'}\Psi_{A_\uparrow(udu)} ^{a'b'c'} (\vec{ x}_1^v,\vec{ x}^v_2,\vec{ x}^v_3)\,.
\end{eqnarray}

To obtain ${\cal N}_p $, we calculate the overlaping
\begin{equation}\label{overlapintegralwithitself}
	\begin{aligned}
	\langle p(\vec{p}\, ') |p( \vec{p}) \, \rangle = {\cal N}^2_p \int& e^{i (\gamma v-\gamma'v') (E_uz_1+E_d z_2 + E_u z_3)}\phi^\dagger_u(\vec{x}_1^{v'} - \vec{x}') S_v^2\phi_u(\vec{x}_1^v - \vec{x})   \\
&
\phi^\dagger_d(\vec{x}_2^{v'} - \vec{x}') S_v^2\phi_d(\vec{x}_2^v - \vec{x}) \phi^\dagger_u(\vec{x}_3^{v'} - \vec{x}') S_v^2 \phi_u(\vec{x}_3^v - \vec{x})  d^3 \vec{x} d^3\vec{x}' [d^3\vec{x}]\,,
	\end{aligned}
\end{equation}
which can be derived from  Eqs.~\eqref{21}, \eqref{24} and \eqref{26}. 
The spin indices have not been written down explicitly as they are irrelevant as long as the inital and final quarks are in the same directions and 
we have taken $S_{v'}^\dagger = S_v$ by anticipating that $v=v'$ or else the integral vanishes.
 To simplify the integral, we adopt the following variables:
\begin{eqnarray}\label{variable_change}
&&\vec{x}_\Delta = \vec{x} - \vec{x}'\,,~~~~\vec{x}_A = \frac{1}{2}(\vec{x}+\vec{x}')\,,~~~~\vec{x}^{ \,r}_i = \vec{x}_{i}^v - \frac{1}{2}(\vec{x}+\vec{x}')\,,
\end{eqnarray}
Now, the  integral is read as 
\begin{eqnarray}\label{overlap}
	&& \frac{{\cal N}^2_p}{\gamma^3}\int d^3\vec{x}_\Delta  d^3 \vec{x}_A\prod_{i=1,2,3}  d^3 \vec{x}^{\,r}_{i} \phi_{q_i}^\dagger\left(\vec{x}^{ \,r}_{i} +\frac{1}{2}\vec{x}_\Delta\right) S^{2}_v \phi_{q_i}\left(\vec{x}^{ \,r}_{i}  -\frac{1}{2}\vec{x}_\Delta\right) e^{iE_{q_i}(v-v')z^r_{i}} e^{iE_{q_i}( v-v'   )z_A  } \nonumber\\
&& = {\cal N}^2 \gamma (2\pi)^3 \delta(\vec{p} -\vec{p}' ) \int d^3\vec{x}_\Delta  \prod_{i=1,2,3}  D^0_{q_i} (\vec{x}_\Delta ) \,,
\end{eqnarray}
where 
$(q_1,q_2,q_3) = (u,d,u)$,  
$1/\gamma^3$ comes from the Jacobian in Eq.~\eqref{variable_change}, and 
$D_{q}^0(\vec{x}_\Delta) $ is defined in Eq.~\eqref{Dvdefin} with $v=0$.
To get the correct $\delta$ function, we have to demand that
\begin{equation}\label{mass}
	M_p=\sum_{i = 1,2,3} E_{q_i}\,,
\end{equation}
for the integral of $\vec{x}_A$, which is the main source of  errors in our model. However,  the range in Eq.~\eqref{energy} shall cover the reasonable values. 
  The $\delta$ function can be interpreted as  that the overlapping of  $D_q^0(\vec{x}_\Delta)$ for the bags with a distance $\vec{x}_\Delta$    occurs  infinite times in the integral, which is essentially a result of the translational symmetry.

By taking the normalization for a momentum state as
\begin{equation}\label{pp'}
	\langle \vec{p},\lambda_p|\vec{p}',\lambda_p' \rangle  =    u^\dagger_p  u_p' (2\pi)^3\delta^3(\vec{p}-\vec{p}')\,,
\end{equation}
where $u_p$ and $\lambda_p$ are  the Dirac spinor and spin of the proton,
we find 
\begin{eqnarray}\label{32}
	\frac{1}{{\cal N}^2_p} &=& \frac{1}{\overline{u}_p u_p }\int d^3\vec{x}_\Delta\prod_{i=1,2,3}   D^0_{q_i} (\vec{x}_\Delta ) \,.
\end{eqnarray}
It is important that ${\cal N}_p$ must be independent of the  velocity because the Lorentz boosts are  unitary for physical states. Here, our result shows that it is indeed the case  in contrast to Eq.~\eqref{wavepacket}\,.
The wave functions and normalization constants of other baryons can be obtained straightforwardly with trivial modifications. In Appendix A, we give the baryon wave functions at rest that are used in this work, and the evaluation of ${\cal N}_p$ can be found in Appendix B.

Here, we summarize few steps we take to construct the hadron wave functions:
\begin{itemize}
\item  We start with the hadrons  at rest, where the translational symmetry is respected.
At this stage, hadrons are not moving yet, so we do not need to worry about the Lorentz symmetry. 
\item The nonzero momentum states are acquired by Lorentz boosts on the physical states, which respect both the translational and  Lorentz symmetries. 
\end{itemize}
Since the Poincar\'e symmetry is preserved in all the steps, we conclude that our approach is consistent with it.  In the next section, we will show explicitly that the form factors do not depend on the Lorentz frame in contrast to the wave packet and ellipsoidal approaches. 

\subsection*{Transition matrix elements}
After the hadron wave functions are constructed, the calculations of the transition matrix elements are straightforward. Here, we  choose  the neutron-proton transition  as an example with $d\to u $  at  quark level. For the calculation, we adopt the Briet frame, where $n$ and $p$ have opposite velocities, and without lost of generality, we take $\vec{v} \parallel \hat{z}$. 
By sandwiching the quark transition operators with the hadron states, we arrive at
\begin{eqnarray}\label{master}
 \langle p (\vec{v}\,), \lambda_p | u ^ \dagger\Gamma  d (0)  |n  (- \vec{v}\,), \lambda_n  \rangle = {\cal N}_{n} {\cal N}_{p} \int d^3\vec{x}_\Delta \Gamma^{\lambda_p\lambda_n} _{ud}(\vec{x}_\Delta) \prod_{q=u,d } D^v_{q}(\vec{x}_\Delta)\,,
\end{eqnarray}
along with
\begin{eqnarray}\label{con}
	&&\Gamma _{ud} ^{\lambda_p\lambda_n} (\vec{ x}_\Delta)=\sum_{\lambda_u,\lambda_d} N^{\lambda_p\lambda_n}_{\lambda_u\lambda_d}\int  d^3\vec{x}  \phi _{u{\lambda_u}}^\dagger\left(\vec{x} ^+ \right) S_{\vec{ v}}\Gamma S_{-\vec{ v}} \phi_{d{\lambda_d}}\left(\vec{x} ^- \right) e^{2i(E_{u} + E_{d})\vec{ v}\cdot \vec{ x}     }\,,
\end{eqnarray}
where  $\Gamma$ is an arbitrary Dirac matrix
and $\vec{x}^\pm = \vec{x}\pm \vec{x}_\Delta /2$.
Here,
$\lambda_{u,d} \in (\uparrow,\downarrow)$ are the spins of the  annihilated up  and down quarks, and 
$N^{\lambda_p\lambda_n}_{\lambda_u\lambda_d}$ represents the overlapping with specific $\lambda_{p,n,u,d}$, which can be computed by matching the LHS and RHS of Eq.~\eqref{master}.
As the values of $N^{\lambda_p\lambda_n}_{\lambda_u\lambda_d}$ is independent of the velocity, it can be obtained by taking $\vec{v} = 0$ for convenient.
From the angular momentum conservation, we have that 
\begin{eqnarray}\label{35}
&&N^{\lambda_p\lambda_n}_{\lambda_u\lambda_d} = 0~~~~~\text{for}~~\lambda_n-\lambda_p \neq \lambda_d - \lambda_u \,,
\end{eqnarray}
It states that if the  baryon spin is (un)flipped by the  operator, then the spin of the quark shall also be  (un)flipped. On the other hand, by the Wiger-Eckart theorem, we have 
\begin{equation}\label{36}
N^{\uparrow \uparrow}_{\lambda_u\lambda_d} =
N^{\downarrow \downarrow}_{-\lambda_u-\lambda_d}\,,~~~~N^{\uparrow \uparrow}_{\uparrow\uparrow} -   N^{\uparrow \uparrow}_{\downarrow\downarrow}= N^{\downarrow\uparrow}_{\downarrow\uparrow} = N^{\uparrow\downarrow}_{\uparrow\downarrow}\,.
\end{equation}
Consequently, there are only two independent numbers in $N^{\lambda_p\lambda_n}_{\lambda_u,\lambda_d}$, given as 
\begin{equation}\label{37}
N_{\text{nonflip}} \equiv  N^{\uparrow\uparrow}_{\uparrow\uparrow} +N^{\uparrow\uparrow}_{\downarrow\downarrow}\,,~~~~N_{\text{flip}} \equiv  N^{\downarrow\uparrow}_{\downarrow\uparrow}\,.
\end{equation}
In the $n\to p$ beta decays, we have $(N_{\text{nonflip}} ,N_{\text{flip}} ) = ( 1, 5/3)$.

Each term  in Eq~\eqref{master} has a  concrete physical meaning, which can be summarized as follows: 
\begin{itemize}
	\item $D_{q}^v(\vec{x}_\Delta) $ are the overlapping  coefficients of the spectator quarks~($u$ and $d$ ) in the initial and final states as found in the ellipsoidal bag approach. Note that  
	the centers of the quark wave functions are separated at a distance of $\vec{x}_\Delta$; 
	\item $\Gamma _{ud} ^{\lambda_p\lambda_n} (\vec{ x}_\Delta)$ is the overlapping coefficient of  $d\to u$ at   quark level. 
	Again, the centers of the bags are separated at a distance of $\vec{x}_\Delta$. 
\end{itemize}
Here, we have found  that the overlapping integrals of the spectator quarks with different velocities~($D^v_{q}$) do not vanish, which is a feature inherited from  the ellipsoidal bag approach. 
However, we have the energy-momentum conservation when we consider the whole wave functions~\cite{Geng:2020ofy}.  It can be viewed as the  spectator quarks are kicked by the bag, which are  in turn  kicked by the quark transition operators.

The main ambiguity of the homogeneous bag model comes from $E_q$ in the exponential as shown in Eq.~\eqref{energy}. However,  
the deviations are insensitive at low velocities
as $E_q$ are always followed by $v$, which make the calculation for the neutron-proton transition unaffected.

For the neutron $\beta$ decay,  the dimensionless form factors $F_1^{V,A}$ are defined by~\cite{Groote:2019rmj}
{\small 
\begin{eqnarray}\label{formfactorss}
&&\langle p (\vec{v}_p) | \overline{u }\gamma^\mu   d (0)  |n  ( \vec{v}_n )  \rangle
=\overline{u}_{p}(\vec{v}_p) \left(
	F^V_1(q^2) \gamma^{\mu} - F^V_2 (q^2)i \sigma^{\mu \nu} \frac{q_\nu}{ M_{n}}   +F^V_3(q^2) \frac{q^{\mu}}{M_{n}}
	\right) u_{n}(\vec{v}_n)\,,
	\\
&&	\langle p (\vec{v}_p)) | \overline{u} \gamma^\mu \gamma^5 d (0)| n (\vec{v}_n)\,) \rangle 
=\overline{u}_{p}(\vec{v}_p) \left(
	F^A_1(q^2) \gamma^\mu - F^A_2 (q^2)i \sigma_{\mu \nu} \frac{q^\nu}{ M_{n}}   +F^A_3(q^2) \frac{q^\mu}{M_{n}}
	\right)\gamma^5 u_{n}(\vec{v}_n)\,,\nonumber
\end{eqnarray}
}
where $q$ corresponds to the 4-momentum difference of the neutron and proton, and $\vec{v}_n$ and $\vec{v}_p$ are the velocities of the neutron and proton, respectively.
 $F_1^V$ and $F_1^A$ can be extracted straightforwardly  after computing the transition matrix elements with $\Gamma= 1$ and $\Gamma = \gamma^0\gamma^{1}\gamma_5$, respectively.
In the model, $u$ and $d$ are taken as masslesss, and thus there is  only one free parameter $R$, having the length dimension. The twist is that $F_1^{V,A}$ do  {\it not}  rely on the bag radius
since the length dimension can not be canceled.
At the $v \to 0 $ limit, we find   that 
\begin{equation}\label{39}
F^V_1 = 1 \,,~~~~F^A_1 = 1.31\,,
\end{equation}
which are close to the experimental value of  $F_1^A/F_1^V = 1.27$\,.
The details of numerical evaluation can be found in Appendix B. 
Compared to the result of  $F_1^A/F_1^V = 1.09$, given previously in the bag model~\cite{DeGrand:1975cf},  the ratio improves significantly after considering the correction from the CMM.

Notice that to compute the matrix elements we have taken the Briet frame, where the initial and  final hadrons have opposite velocities. However, in principle, it can be calculated in  other Lorentz frames also. For an illustration, with $\Gamma = \gamma^0 \gamma^\mu$, we have 
\begin{eqnarray}
	&& \langle p ({v}\,), \lambda_p | \overline{u }\gamma^\mu   d (0)  |n  (- \vec{v}\,), \lambda_n  \rangle
	=\langle p (0), \lambda_p | U_{-v}\overline{u}\gamma^\mu   d (0) U_{-v} |n  (0), \lambda_n  \rangle\nonumber\\
	&&= \langle p (0), \lambda_p | U^2_{-v}U_v\overline{u}\gamma^\mu  d (0) U_{-v} |n  (0), \lambda_n  \rangle =
	\langle p (\vec{v}\,'), \lambda_p |\overline{u} S_{v}\gamma^\mu S_{-v} d (0)  |n  (0), \lambda_n  \rangle \nonumber\\
	&&= \langle p (\vec{v}\,'), \lambda_p |\overline{u} (\Lambda_{-v})^\mu\,  _\nu \gamma^\nu d (0)  |n  (0), \lambda_n  \rangle  = (\Lambda_{-v})^\mu\,  _\nu \langle p (\vec{v}\,'), \lambda_p |\overline{u}  \gamma^\nu d (0)  |n  (0), \lambda_n  \rangle \,,
\end{eqnarray}
where $U_v^2 = U_{v'}$,
the use of Eq.~\eqref{Lorentz} has been made in the  second line, and $S_{-v} \gamma^{\mu'} S_{v} = (\Lambda_v)^{\mu'}\,_{\nu}\gamma^\nu$ in the third line. 
Plugging in the last equation in Eq.~\eqref{formfactorss}, we find that 
\begin{eqnarray}
&&\langle p(\vec{v}\,')  | \overline{u} \gamma^\mu d(0) | n(0)\rangle 
= (\Lambda^{-1}_v)^\mu\,_{\mu'} \overline{u}_{p}(\vec{v}\,') \left(
F^V_1(q^2) \gamma^{\mu'} - F^V_2 (q^2)i \sigma^{\mu' \nu} \frac{q'_\nu}{ M_{n}}   +F^V_3(q^2) \frac{q'^{\mu'}}{M_{n}}
\right)u_{n}(0)\nonumber\\
&&=(\Lambda^{-1}_{v})^{\mu}\,_{\mu '} \overline{u}_{p}(\vec{v}\,)S_{-v} \left(
F^V_1(q^2) \gamma^{\mu'} - F^V_2 (q^2)i \sigma^{\mu' \nu} \frac{q'_\nu}{ M_{n}}   +F^V_3(q^2) \frac{q'^{\mu'}}{M_{n}}
\right)S_v u_{n}(-\vec{v})\nonumber\\
&&= \overline{u}_{p}(\vec{v}\,) \left(
F^V_1(q^2) \gamma^{\mu} - F^V_2 (q^2)i \sigma^{\mu \nu} \frac{q_\nu}{ M_{n}}   +F^V_3(q^2) \frac{q^{\mu}}{M_{n}}
\right) u_{n}(-\vec{v})\,,
\end{eqnarray}
with $q'^{\mu} = (\Lambda_v)^\mu\,_\nu q^\nu $, 
which are identical to 
\begin{equation}
\langle p(\vec{v})  | \overline{u} \gamma^\mu d(0) | n(-\vec{v})\rangle 
= \overline{u}_{p}(\vec{v} ) \left(
F^V_1(q^2) \gamma^{\mu} - F^V_2 (q^2)i \sigma^{\mu \nu} \frac{q_\nu}{ M_{n}}   +F^V_3(q^2) \frac{q^{\mu}}{M_{n}}
\right)u_{n}(-\vec{v})\,.
\end{equation}
Thus,  the results are independent of the Lorentz frame we choose in contrast to the ellipsoidal bag approach. 
Combing the feature of the wave functions to be invariant under spacetime translations, we conclude that
 the Poincar\'e symmetry is recovered.

For the heavy quark transitions, we use the decay of 
 $\Lambda_b \to \Lambda \gamma$ 
for an example, which is governed by  the tensor operator of $b\to s$.
The bag radii of $\Lambda$ and $\Lambda_b$ are found to be around $5$ and $4.6$~GeV$^{-1}$, respectively~\cite{Hypothetic3,DeGrand:1975cf}.
We take both of them as $4.8$~GeV$^{-1}$ to simplify the numerical calculations. In addition, we find that the results depend little on the quark masses as long as the values are reasonable. For simplicity, we use~\cite{ pdg}
\begin{equation}
(M_s , M_b) = (0.1, 4.78)~\text{GeV}\,,
\end{equation} 
where $M_s$ and $M_{b}$ are  taken as  the current  and pole masses, respectively. 
The  tensor form factors are defined as 
\begin{equation}
\langle \Lambda | \overline{s} i \sigma^{\mu\nu} q_\nu \gamma_5 b| \Lambda_b\rangle 
=\overline{u}_\Lambda \left[
f_1^{T A}\left(q^2\right)\left(\gamma^\mu q^2-q^\mu q_\nu \gamma^\nu  \right) / M_{\Lambda_b}-f_2^{T A}\left(q^2\right) i \sigma^{\mu \nu}q_\nu 
\right]\gamma_5
\overline{u}_{\Lambda _b}\,,
\end{equation}
of which 
only $f_2^{TA}$ is relevant
to  the weak radiative decay.  It can be calculated by taking  $\Gamma  = \gamma^0 \sigma^{1\nu} q_{\nu} $ in Eq.~\eqref{master} with a slight modification, read as 
\begin{equation}
 \langle \Lambda  (\vec{v}\,), \lambda_{\Lambda} | s ^ \dagger  \Gamma b (0)  |\Lambda_b  (- \vec{v}\,), \lambda_{\Lambda_b} \rangle = {\cal N}_{\Lambda} {\cal N}_{\Lambda_b} \int d^3\vec{x}_\Delta \Gamma^{\lambda_{\Lambda}\lambda_{\Lambda_b}} _{\lambda_s\lambda_b}(\vec{x}_\Delta) \prod_{q=u,d } D^v_{q}(\vec{x}_\Delta)\,,
\end{equation}
with 
\begin{eqnarray}\label{con2}
\Gamma^{\lambda_{\Lambda}\lambda_{\Lambda_b}} _{\lambda_s\lambda_b}(\vec{ x}_\Delta)=\sum_{\lambda_s,\lambda_b} N^{\lambda_{\Lambda}\lambda_{\Lambda_b}} _{\lambda_s\lambda_b}\int  d^3\vec{x}  \phi _{s{\lambda_s}}^\dagger\left(\vec{x} ^+ \right) S_{\vec{ v}}\Gamma S_{-\vec{ v}} \phi_{b{\lambda_b}}\left(\vec{x} ^- \right) e^{2i(E_{u} + E_{d})\vec{ v}\cdot \vec{ x}     }\,.
\end{eqnarray} 
For $E_u$ in the range of Eq.~\eqref{energyn},   the form factor is found to be
\begin{equation}
f_2^{TV}\left(q^2=0\right) = 0.13 4\pm 0.034 \,,
\end{equation}
leading to 
\begin{equation}
{\cal B} (\Lambda_b \to \Lambda \gamma) =  ( 6.8 \pm 3.3 ) \times 10 ^{ -6} \,,
\end{equation} 
where  the numerical evaluations of the form factors are given in Appendix B. The formalism of the branching fraction can be founded in   Ref.~\cite{addedd}, given as 
\begin{equation}
{\cal B}\left(\Lambda_b \rightarrow \Lambda \gamma\right)=\frac{\tau_{b }\alpha_{e m}}{32 \pi^4} G_F^2 M_b^2 M_{\Lambda_b}^3\left|V_{t s} ^* V_{t b}\right|^2\left(C_{7 \gamma}^{e f f}\right)^2\left(1-\frac{M_{\Lambda}^2}{M_{\Lambda_b}^2}\right)^3 \left|f_2^{T A}\right|^2\,,
\end{equation}
where $\tau_b$ is the lifetime of $\Lambda_b$, 
$G_F$ is the Fermi constant, $\alpha_{e m} = 1/137$, $C_{7 \gamma}^{e f f} =0.303$, and $M_b = 4.8$ GeV. 
Our result of  the branching ratio is  consistent with  $(7.1\pm 1.7)\times 10^{-6}$ given by the experiment~\cite{pdg}.  In contrast to Eq.~\eqref{39}, the form factors of $\Lambda_b \to \Lambda$ suffer large uncertainties from the quark energies since the Lorentz boost with high velocity~\footnote{To be specific, the velocity is found to be 0.669. } is needed.   
Alternatively, one can fit the quark energies from the experiments, resulting in 
\begin{equation}
E_{u,d}= (0.33\pm 0.01) ~\text{GeV}\,,
\end{equation}
 which is consistent with  Eq.~\eqref{energyn} and useful for the future work.

\section{Summary}
We have reviewed  the attempts 
tackling with the CMM of the bag model in the literature. 
We have discussed the advantages of the wave packet and the ellipsoidal bag approaches as well as their inconsistencies with the Poincar\'e symmetry.
By combing their merits, we have proposed the framework of the homogeneous bags, which is consistent with the Poincar\'e symmetry. 
Notably, we have shown that in our framework, the dominated form factors of the neutron $\beta$ decay do not depend on any free parameters, given as  $F_1^A/F_1^V = 1.31$,  which is  close to the experimental value of  $1.27$. 
For the heavy quark transition, we have taken the decay of  $\Lambda_b \to \Lambda \gamma$ as an example,  and obtained that  ${\cal B}(\Lambda_b \to \Lambda \gamma) = (6.8 \pm 3.3) \times 10^{-6} $, which is consistent with the experimental measurement of $(7.1 \pm 1.7 ) \times 10^{-6}$. 
As a conclusion, we have found that the homogeneous bag model is  useful in both   light and heavy   quark systems. 
It is clear that  the homogeneous bag model can provide a reliable 
framework for the computations  concerning the hadron transitions, including the  form factors as well as the decay constants.

\appendix
\section{Baryon wave functions at rest}
We collect the baryon wave functions at rest that are used in this work:
\begin{eqnarray}\label{wavefunctionsAPP}
&&|p,\uparrow \rangle =  \int [d^3\vec{x}] 
\frac{1}{2 \sqrt{3}}
\epsilon^{\alpha \beta \gamma} u^\dagger_{a\alpha}(\vec{x}_1)d^\dagger_{b\beta}(\vec{x}_2)u^\dagger_{c\gamma}(\vec{x}_3)	\Psi_{A_\uparrow (udu)}^{abc} ( \vec{x}_1, \vec{x}_2 ,  \vec{x}_3 )|0\rangle \,,\nonumber\\
&&|n ,\uparrow\rangle =  \int [d^3\vec{x}] 
\frac{1}{2 \sqrt{3}}
\epsilon^{\alpha \beta \gamma} u^\dagger_{a\alpha}(\vec{x}_1)d^\dagger_{b\beta}(\vec{x}_2)d^\dagger_{c\gamma}(\vec{x}_3)	\Psi_{A_\uparrow (udd)}^{abc} ( \vec{x}_1, \vec{x}_2 ,  \vec{x}_3 )|0\rangle \,,\nonumber\\
&&|\Lambda,\uparrow\rangle =  \int [d^3\vec{x}] 
\frac{1}{ \sqrt{6}}
\epsilon^{\alpha \beta \gamma} u^\dagger_{a\alpha}(\vec{x}_1)d^\dagger_{b\beta}(\vec{x}_2)s^\dagger_{c\gamma}(\vec{x}_3)	\Psi_{A_\uparrow (uds)}^{abc} ( \vec{x}_1, \vec{x}_2 ,  \vec{x}_3 )|0\rangle \,,\nonumber\\
&&|\Lambda_b ,\uparrow \rangle =  \int [d^3\vec{x}] 
\frac{1}{\sqrt{6}}
\epsilon^{\alpha \beta \gamma} u^\dagger_{a\alpha}(\vec{x}_1)d^\dagger_{b\beta}(\vec{x}_2)b^\dagger_{c\gamma}(\vec{x}_3)	\Psi_{A_\uparrow (udb)}^{abc} ( \vec{x}_1, \vec{x}_2 ,  \vec{x}_3 )|0\rangle \,.
\end{eqnarray}

\section{Details of evaluating the form factors}
The calculations of the form factors are  tedious works.
In principle, one can evaluate Eqs.~\eqref{master} and \eqref{con} numerically and  match them with the form factors defined in Eq.~\eqref{formfactorss}. However, in practice there are too many integrals, and we have to carry out some of the angular ones  so that the matrix elements can be evaluated numerically at a reasonable time by a computer program.

\begin{figure}
\centering
\includegraphics[width=0.4\linewidth]{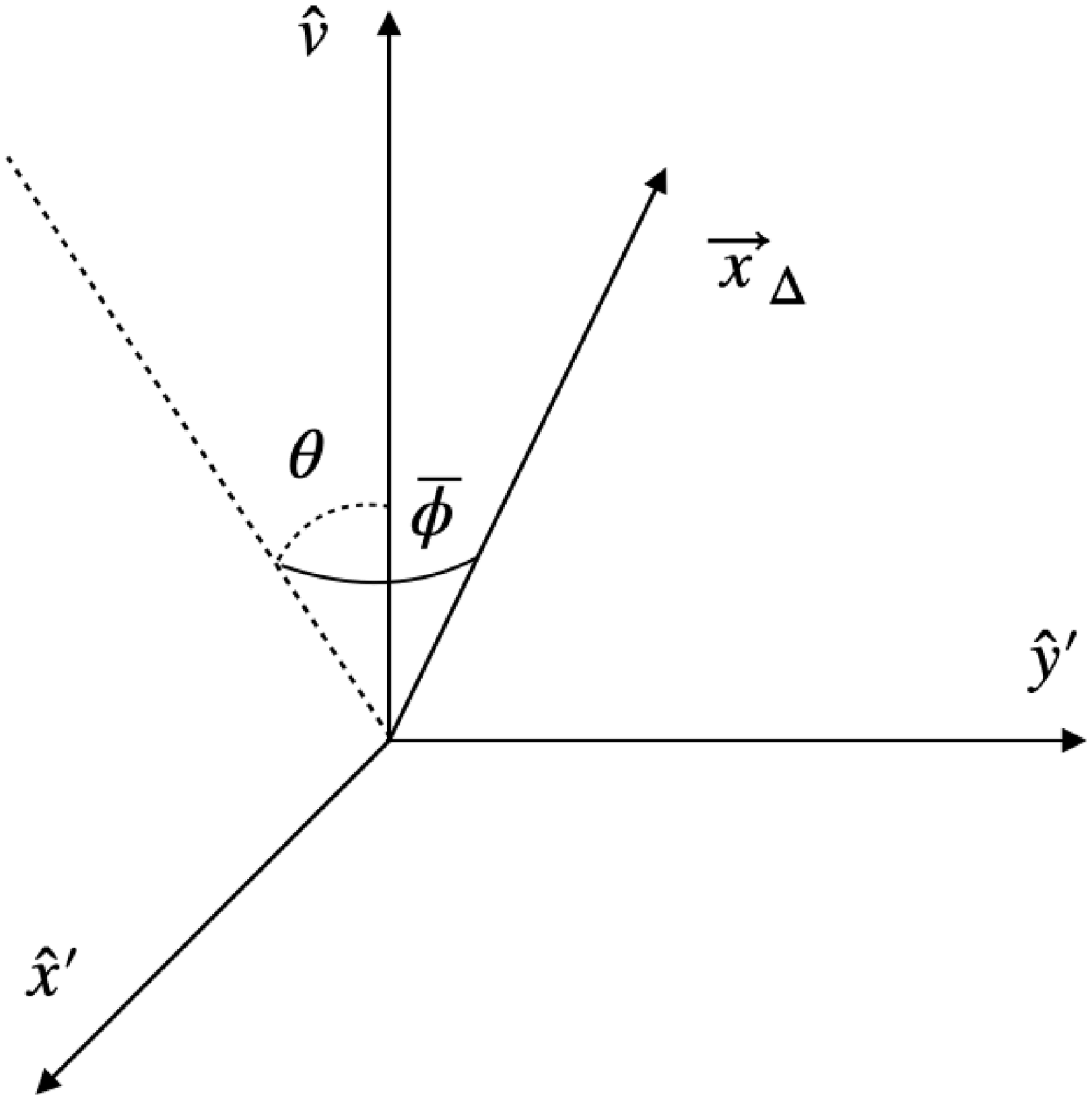}
\includegraphics[width=0.47\linewidth]{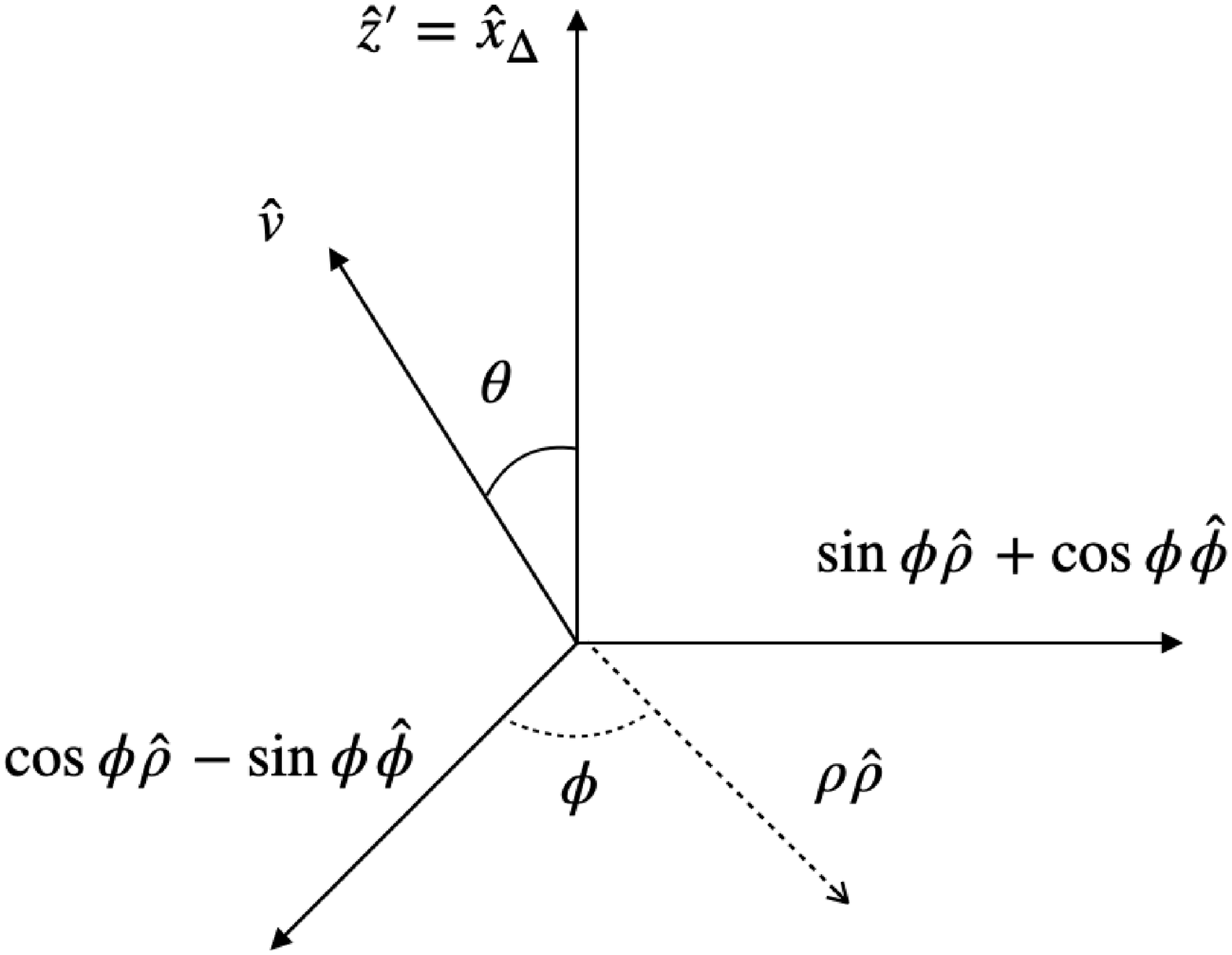}
\caption{The definitions of the angles $\theta, \overline{\phi}$ and $\phi$, where the right figure is  the  adopted cylindrical coordinate in evaluating ${\cal D}^v_q(\vec{x}_\Delta)$ . }
\label{FIG1}
\end{figure}

One of the important observations is that   $D^v_{q}(\vec{x}_\Delta)$ defined in Eq.\eqref{Dvdefin} 
does not depend on the quark spin.  There are only two  directions~($\vec{ v}$ and $\vec{x}_\Delta$) that are specified in the integral.
 Therefore, $D^v_{q}(\vec{x}_\Delta)$ can only depend on their magnitudes and product, read as 
\begin{equation}\label{B1}
D^v_{q}( \vec{x}_\Delta ) = D^v_{q}( r_\Delta , \cos \theta)\,,
\end{equation}
where $r_\Delta= |\vec{x}_\Delta|$ and 
$\theta$ is the angle between $\vec{x}_\Delta$ and $\vec{v}$. 
Accordingly, we can rotate both $\vec{v}$ and $\vec{x}_\Delta$ simultaneously without affecting the numerical results. We  adopt the cylindrical coordinate~$(\rho,\phi,z')$ and use the freedom to choose $\vec{x}_\Delta\parallel \vec{z}\,'\,
$ with $\vec{v}$ lying on the $\hat{ \rho}\otimes \hat{z}'$ plane at  $\phi = 0$. To be specific, we take 
\begin{equation}\label{B2}
\vec{x} = \rho \hat \rho + z' \hat{ z}'\,,~~~~
\vec{x}_\Delta = r_\Delta \hat{z}'\,,~~~~\vec{v} = v \left(
\sin \theta \cos \phi \hat{\rho}  - \sin \theta \sin \phi \hat{\phi}  + \cos \theta \hat z '
\right)\,,
\end{equation}
where $\vec{x}$ is the integration variable in Eq.~\eqref{Dvdefin}, and 
the definitions of the angles are collected in FIG.~\ref{FIG1}.
Plugging Eq.~\eqref{B2} in  Eq.~\eqref{Dvdefin}, we arrive at 
\begin{eqnarray}\label{specified}
&&	\gamma D^v_{q}(\vec{x}_{\Delta}) = \int d^3 \vec{x} \phi_{q}^\dagger \left(\vec{x}^+\right) \phi_{q} \left(\vec{x}^-\right)
	e^{-2iE_{q}  \vec{ v}\cdot\vec{ x} }=\int d^3 \vec{x} \phi_{q}^\dagger \left(\vec{x}^+\right) \phi_{q} \left(\vec{x}^-\right)
e^{-2iE_{q} v (\sin\theta \rho \cos \phi + \cos\theta z')} \nonumber\\
	&&=
	\int \rho d\rho d\phi dz '
	\left[
	j_{0q}^+j_{0q}^- + j_{1q}^+j_{1q}^-\left( 
	\hat{x}^+ \cdot \hat{x}^- +\frac{1}{r^+r^- } i  \chi^\dagger  (\vec{x}_\Delta \times \vec{x}) \cdot \vec{\sigma}\chi
	\right)
	\right]e^{-2iE_{q} v (\sin\theta \rho \cos \phi + \cos\theta z')}\nonumber\\
	&&=
\int \rho d\rho d\phi dz '
\left(
j_{0q}^+j_{0q}^- + j_{1q}^+j_{1q}^- 
\hat{x}^+ \cdot \hat{x}^- 
\right)e^{-2iE_{q} v (\sin\theta \rho \cos \phi +  \cos\theta z')}\nonumber\\
	&&=2\pi  \int \rho d\rho  dz  '
	\left(
	j_{0q}^+j_{0q}^- +  j_{1q}^+j_{1q}^- \hat{x^+} \cdot  \hat{x^-} 
	\right)J_0(2E_{q}v\sin \theta \rho)\cos \left( 2 E_{q}   v \cos\theta z' \right) \,,
\end{eqnarray}
where $J_0$ is the zeroth Bessel function, 
\begin{equation}
	j_{(0,1)q}^{\pm} \equiv  \omega_{q(+,-)} j_{(0,1)} ( p_{q} r^\pm) \,,
\end{equation}
with $r^\pm = |\vec{x}^\pm|$. Here, 
 we have absorbed $N$ to the overall normalization constant ${\cal N}_{n,p}$ for convenience. 
Due to the finite bag radius, the integrals are bounded as 
\begin{equation}
\int \rho d\rho  d z'  =\int_0^{ \sqrt{R^2-r_\Delta^2/4}} d \rho \rho  \int_{ - \sqrt{R^2-r_\Delta^2/4} + r_\Delta/2 }^{ \sqrt{R^2-r_\Delta^2/4} - r_\Delta/2 }  d z' \,.
\end{equation}
For the sake of compactness, the regions of the integrations for $\rho$ and $z'$ are not written down explicitly as long as  there is no confusion.
We have dropped 
$\chi^\dagger[ (\vec{x}_\Delta \times \vec{x})\cdot \vec{\sigma}] \chi$\,,
since after integrating $d^3\vec{x}$, it  is proportional to $\chi^\dagger[( \vec{x}_\Delta\times \hat{v})\cdot \vec{\sigma}] \chi$ as $\vec{ v}$ is the only specified direction. It vanishes since we always choose the spin  directions at  $\pm \hat{v}$. 
In the last line of Eq.~\eqref{specified}, we have used 
\begin{equation}
J_n(a)  = \frac{1}{2\pi }
\int^{\pi}_{-\pi } \exp( i n \phi- ia \sin \phi  ) d \phi \,,
\end{equation}
with the integrand being an even function of  $z'$.  Notice that Eq.~\eqref{specified} is consistent with Eq.~\eqref{B1} due to that $J_0(a) = J_0(-a)$, which implies that $J_0(2 E_{q} v \sin \theta \rho)=J_0(2 E_{q} v \sqrt{1 - \cos^2 \theta}  \rho)$. 
To conclude,  the number of integrals in Eq.~\eqref{Dvdefin} is reduced as two, which greatly shorten the evaluating time. 

To compute the normalization constant,
we take $v=0$ in Eq.~\eqref{specified} and arrive at 
\begin{eqnarray}
D_{q}^0 (r_\Delta)&=&2\pi  \int \rho d\rho  dz  '
\left(
j_{0q}^+j_{0q}^- +  j_{1q}^+j_{1q}^- \hat{x^+} \cdot  \hat{x^-} 
\right)\,,\nonumber\\
\frac{1}{{\cal N}_{n,p}^2} &=& \frac{1}{\overline{u}_{n,p} u _{n,p}}\int 4\pi r_\Delta^2 d r_\Delta  \left( D_u^0(r_\Delta)\right) ^3 \,,
\end{eqnarray}
where we employ the isospin symmetry $D_u^v = D_d^v$. 

Now we turn our attention to  $\Gamma_{ud}^{\lambda_p\lambda_n}(\vec{x}_\Delta)$. 
From Eq.~\eqref{specified}, we find that $D^v_{q}(\vec{x}_\Delta)$ is an even function of $\vec{x}_\Delta$. Thus, we can drop the terms that are odd regarding to $\vec{x}_\Delta$ in $\Gamma_{ud}^{\lambda_p\lambda_n}(\vec{x}_\Delta)$. In this work, we use $\Gamma = 1 $ and $\Gamma = \gamma^0 \gamma^1\gamma_5$ as  examples. To evaluate  $\Gamma=1$, we take $\lambda_n =\lambda_p = \uparrow$ in Eqs.~\eqref{master} and \eqref{con}, resulting in
\begin{eqnarray}
&&\begin{aligned}
\Gamma _{ud} ^{\uparrow \uparrow } (\vec{ x}_\Delta)= N_{\uparrow\uparrow}^{\uparrow\uparrow}&\int  d^3\vec{x}  \phi _{u{\uparrow}}^\dagger\left(\vec{x} ^+ \right)  \phi_{d{\uparrow}}\left(\vec{x} ^- \right) e^{2iE_{\text{di}}\vec{ v}\cdot \vec{ x}     } +
N_{\downarrow\downarrow}^{\uparrow\uparrow}\int  d^3\vec{x}  \phi _{u{\downarrow}}^\dagger\left(\vec{x} ^+ \right)  \phi_{d{\downarrow}}\left(\vec{x} ^- \right) e^{2iE_{\text{di}}\vec{ v}\cdot \vec{ x}     } 
\end{aligned}\nonumber\\
&&\quad= N_{\text{nonflip}}
\int  d^3\vec{x}  \phi _{u{\uparrow}}^\dagger\left(\vec{x} ^+ \right)  \phi_{d{\uparrow}}\left(\vec{x} ^- \right) e^{2iE_{\text{di}}\vec{ v}\cdot \vec{ x}     } \nonumber\\
&&\quad=2\pi  \int \rho d\rho  dz  '
\left(
j_{0u}^+j_{0q_d}^- +  j_{1q_u}^+j_{1q_d}^- \hat{x^+} \cdot  \hat{x^-} 
\right)J_0(\delta_\rho)\cos \left( \delta_z \right) \,,
\nonumber\\
&&\delta_\rho \equiv  2 E_{\text{di}} v \sin \theta \rho \,,~~~~~\delta_z \equiv 2 E_{\text{di}} v \cos \theta z' 
\end{eqnarray}
where $E_{\text{di}}= E_{u} + E_{d}$ is the energy of the spectator quarks. Finally, we obtain
\begin{eqnarray}\label{B7}
&&\langle p (\vec{v}\,), \uparrow | u ^ \dagger  d (0)  |n  (- \vec{v}\,), \uparrow  \rangle \nonumber\\
&&\quad\quad= {\cal N}_{n} {\cal N}_{p}  2\pi \int^{2R}_0  r_\Delta^2 dr_\Delta \int^{1}_{-1} d\cos \theta \Gamma^{\uparrow\uparrow} _{ud}(r_\Delta,\cos\theta) \left( D^v_{u}(r_\Delta,\cos \theta) \right) ^2 \,.
\end{eqnarray}
In the neutron $\beta$ decay,  we can safely set $v\to 0 $ and  neglect the contributions from $F_{2,3}^{V,A}$.  Comparing the right hand sides of Eqs.~\eqref{formfactorss} and \eqref{B7}, we find $F_1^V = 1 $. 

Now we turn our attention to 
  $\Gamma = \gamma^0 \gamma^1 \gamma_5$. The trick of Eq.~\eqref{B1} can not be applied as $\Gamma$  provides an extra direction.
In the cylindrical coordinate described in 
Eq.~\eqref{B2}, we have 
\begin{eqnarray}\label{xhat}
&&\Gamma  = \gamma^0 \gamma^1 \gamma_5 =\left(
\begin{array}{cc}
\hat{ x }' \cdot \vec{\sigma}& 0 \\
0 & \hat{ x}'\cdot\vec{\sigma}
\end{array}
\right)\,,\\
&&\hat{x}' =
\left(-
\sin \overline{\phi} \sin \phi - \cos \overline{\phi} \cos \theta \cos \phi
\right)\hat{\rho} +\left(
\cos \overline{\phi} \cos \theta \sin \phi - \sin \overline{\phi}\cos 
\phi
\right)+
 \sin\theta \cos \overline{\phi} \hat{z}'\nonumber\,,
\end{eqnarray}
where $\overline{\phi}$ is the azimuthal angle between the $\vec{v} \otimes \vec{x}'$ and $\vec{ v}\otimes \vec{x}_\Delta$ planes. In addition, we have 
\begin{eqnarray}\label{yhat}
&&\hat{x}'\cdot \vec{\sigma} \hat{v} \cdot\vec{\sigma} = -i\hat{y}' \cdot 
\vec{\sigma}\,,~~~~~\hat{y}'\cdot \vec{\sigma} \hat{v}\cdot\vec{\sigma} = i \hat{x}'\cdot\vec{\sigma}\\
&&\hat{y}' =
\left(
\cos \overline{\phi} \sin \phi - \sin \overline{\phi} \cos \theta \cos \phi
\right)\hat{\rho} +\left(
\sin \overline{\phi} \cos \theta \sin \phi + \cos \overline{\phi}\cos 
\phi
\right)+
\sin\theta \sin \overline{\phi} \hat{z}'\nonumber\,.
\end{eqnarray}
Here, $\hat{x}'$ and $\hat{y}'$ point toward the $x$ and $y$ directions, respectively, when we choose $\vec{ v}\parallel \hat{z}$ with Cartesian coordinate system~(see FIG.~\ref{FIG1}). 
We define 
\begin{equation}
{\cal G} \equiv 
\gamma^0\gamma^1 \gamma_5 S_{-v}^2 = \left(
\begin{array}{cc}
	\hat{ x }' \cdot \vec{\sigma}& 0 \\
	0 & \hat{ x}'\cdot\vec{\sigma}
\end{array}
\right)
\left(
\begin{array}{cc}
\gamma & -\gamma \vec{v}\cdot \vec{\sigma} \\
-\gamma \vec{v}\cdot \vec{\sigma} &\gamma 
\end{array}
\right)=\gamma \left(
\begin{array}{cc}
	\hat{ x }' \cdot \vec{\sigma}& iv \hat{y}'\cdot \vec{\sigma} \\
iv \hat{y}'\cdot \vec{\sigma} & \hat{ x}'\cdot\vec{\sigma}
\end{array}
\right)\,.
\end{equation}
To calculate Eq.~\eqref{con}, we choose $(\lambda_n,\lambda_p) =( \uparrow, \downarrow)$, resulting in that 
\begin{eqnarray}\label{B13}
\Gamma _{ud} ^{\downarrow \uparrow} (\vec{ x}_\Delta) &=& {\cal N}_{\text{flip}}\int  d^3\vec{x} {\cal I}(\vec{x}_\Delta) e^{2i(E_{u} + E_{d})\vec{ v}\cdot \vec{ x}     }\,,\nonumber\\
{\cal I} (\vec{x}_\Delta)  &\equiv&   \phi _{u{\downarrow}}^\dagger\left(\vec{x} ^+ \right) {\cal G} \phi_{d{\uparrow}}\left(\vec{x} ^- \right) \nonumber\\
&=&  \left(
\begin{array}{cc}
j_{0u}^+ \chi ^\dagger_\downarrow & -ij_{1u}^+ \hat{x}^+\cdot\vec{\sigma}\chi ^\dagger_\downarrow 
\end{array}
\right)
{\cal G}
\left(
\begin{array}{c}
j_{0d}^- \chi_\uparrow\\
ij_{1d}^- \hat{x}^-\cdot\vec{\sigma}\chi _\uparrow 
\end{array}
\right)\,,
\end{eqnarray}
where we have used 
Eqs.~\eqref{35} and \eqref{37} along with
$S_v \gamma^0\gamma^ 1 \gamma_5 = \gamma^0\gamma^ 1 \gamma_5
S_{-v}$. 
The integrand ${\cal I}$ can be further simplified by  noting 
\begin{eqnarray}
&&\chi^\dagger_\downarrow \chi_\uparrow 
=0\,,~~~~
\chi_\downarrow ^\dagger \vec{\sigma }\chi_\uparrow = \hat{x}' + i \hat{y}'\,,\nonumber\\
&&\sigma_i\sigma_j\sigma_k = i \epsilon_{ijk} + \delta_{jk} \sigma_i - \sigma_j \delta_{ik} + \delta_{ij} \sigma_k \,,
\end{eqnarray}
leading to 
\begin{eqnarray}\label{I def}
&&\frac{1}{2}\left( {\cal I}(\vec{x}_\Delta) + {\cal I }(-\vec{x}_\Delta) \right) = {\gamma }
({\cal I}_1 + v {\cal I}_2 +{\cal I}_3)
\,,\\
&& {\cal I}_1 = {\cal J}_{00}\,,~~~~~{\cal I}_2 =- i\left[{\cal J}_{01} \hat{v}\cdot \hat{x}^- +{\cal J}_{10} \hat{v}\cdot \hat{x}^+\right]\,,\nonumber\\
&&{\cal I}_3 = {\cal J}_{11} \left(
2 \hat{x}^+\cdot\hat{x}' \hat{x}^- \cdot \hat{x}'  + i \hat{x}^+ \cdot\hat{y}' \hat{x}^- \cdot\hat{x}' 
+i \hat{x}^-  \cdot\hat{y}' \hat{x}^+ \cdot\hat{x}'  - \hat{x}^+ \cdot\hat{x}^-
\right)\,,\nonumber
\end{eqnarray}
where 
$\chi_\uparrow$ and $\chi_\downarrow$ stand for  the quark spins pointing toward the  $\hat{v}$ and $-\hat{v}$ directions, respectively, and the first line of Eq.~\eqref{I def} is due to  that we only consider the even part of the integrand regarding to $\vec{x}_\Delta$.  
For the sake of compactness, we have defined 
\begin{equation}
{\cal J}_{nm} \equiv \frac{1}{2}\left( j_{n u}^+ j_{m d}^- + j_{m u}^- j_{n d}^+\right)~~~~\text{for}~~n,m\in \{ 0, 1 \} \,.
\end{equation}
In the cylindrical coordinates described in Eq.~\eqref{B2}, ${\cal J}_{nm}$ depends on $\rho\,,~ z'$ and $r_\Delta$ only, with the following property
\begin{equation}
{\cal J}_{nm}(\rho , - z' ,r_\Delta ) = {\cal J}_{mn}(\rho , z' ,r_\Delta )\,.
\end{equation}
Accordingly, we find that ${\cal J}_{00}$ and ${\cal J}_{11}$ are even functions of   $z'$, whereas ${\cal J}_{01}/r^+ \pm 
{\cal J}_{10}/r^-$ are even and odd, respectively.

With Eq.~\eqref{B2}, 
the integrals  of ${\cal I}_1$ and ${\cal I}_2$ can be  computed  straightforwardly similar to Eq.~\eqref{specified}, given as 
\begin{eqnarray}\label{B16}
	&&\int  d^3\vec{x} {\cal I}_1 e^{2iE_{\text{di}}\vec{ v}\cdot \vec{ x}     }  =2\pi   \int \rho d\rho  dz  '
{\cal J}_{00} J_0(\delta_\rho)\cos \left( \delta_ z \right) \,,\\
&& \int d^3 \vec{x} {\cal I}_2 e^{2iE_{\text{di}}\vec{ v}\cdot \vec{ x}     } 
=2\pi \int \rho d\rho dz'  \Bigg[ \left(   \frac{{\cal J}_{10}}{r^+} - \frac{{\cal J}_{01}}{r^-}  \right)\frac{r_\Delta}{2} \cos \theta  J_0 (\delta_\rho)  \sin (\delta_z) \nonumber\\
&&\qquad\qquad+ \left(  \frac{{\cal J}_{10}}{r^+}+ \frac{{\cal J}_{01}}{r^-}  \right)\Big( \cos \theta z  J_0(\delta_\rho)
\sin \left( \delta_z  \right)  
+ \sin \theta \rho J_1(\delta_\rho)
\cos \left( \delta_z \right)
\Big)\Bigg] 
\,. \nonumber
\end{eqnarray}
On the other hand, 
from Eq.~\eqref{I def}, we see that ${\cal I}_3$ depends also on the azimuthal angle  $\overline{\phi}$. 
To compute
\begin{equation}
 \int dr_\Delta  d \cos \theta d \overline{\phi}\int  d^3\vec{x} {\cal I}_3 e^{2iE_{\text{di}}\vec{ v}\cdot \vec{ x}     }\prod_{q=u,d } D^v_{q}(r_\Delta, \cos \theta)\,,
\end{equation}
we interchange the order of the integrals of $\int d\overline{\phi}$ and $\int d^3\vec{x}$. In addition, we make use of  that $D_{q}^v$ are independent of $\overline{\phi}$, leading to 
\begin{eqnarray}
&&\int  d \overline{\phi}\int  d^3\vec{x} {\cal I}_3 e^{2iE_{\text{di}}\vec{ v}\cdot \vec{ x}     }\prod_{q=u,d } D^v_{q}(r_\Delta, \cos \theta) \nonumber\\
&&\qquad\qquad = 
\prod_{q_j=u,d } D^v_{q_j}(r_\Delta, \cos \theta) \int  d^3\vec{x} \int  d \overline{\phi} {\cal I}_3 e^{2iE_{\text{di}}\vec{ v}\cdot \vec{ x}     }\,.
\end{eqnarray}
Therefore, we can first calculate the integrals of the azimuthal angles.
By explicit calculations, we find 
\begin{eqnarray}\label{explicit}
&&\int d\phi \int d\overline{\phi} 2 \hat{x}^+\cdot\hat{x}' \hat{x}^- \cdot \hat{x}' e^{2iE_{\text{di}}\vec{ v}\cdot \vec{ x}     } =\frac{4\pi^2 }{r^+r^-}\bigg[
\rho^2 \cos^2 \theta J_0(\delta_\rho) \cos(\delta_z) + \frac{1}{2}\rho^2 \sin^2 \theta \big(
J_0(\delta_\rho) 
\nonumber\\
&&\qquad+ J_2(\delta_\rho) 
\big) \cos(\delta_z)+\sin^2 \theta \left(
z^2 - \frac{r_\Delta^2}{4} 
\right)J_0 (\delta_\rho) \cos (\delta_z) + \sin(2 \theta)\rho z J_1(\delta_\rho) \sin(\delta_z) \bigg]\,,\nonumber\\
&&\int d\phi \int d\overline{\phi} \left(  \hat{x}^+\cdot\hat{x}' \hat{x}^- \cdot \hat{y}' +  \hat{x}^+\cdot\hat{y}' \hat{x}^- \cdot \hat{x}'
\right)  e^{2iE_{\text{di}}\vec{ v}\cdot \vec{ x}     }  = 0 \,,
\nonumber\\
&&\int d\phi \int d\overline{\phi}   \hat{x}^+\cdot \hat{x}^-  e^{2iE_{\text{di}}\vec{ v}\cdot \vec{ x}     }   = \frac{4\pi^2 }{r^+r^-}\left(
\rho^2 + z^2 - \frac{r_\Delta^2}{4}
\right) J_0(\delta_\rho) \cos (\delta_z )\,.
\end{eqnarray}
We define 
\begin{equation}
{\cal I}_3' \equiv \frac{1}{2\pi} \int {\cal I}_3 d \overline{\phi}\,,
\end{equation}
of which ${\cal I}_3'$ is independent of $\overline{\phi}$. 
Effectively, one can substitute ${\cal I}_3 '$ for ${\cal I}_3$ without affecting the numerical results. Collecting  Eqs.~\eqref{I def} and \eqref{explicit}, we arrive at
\begin{eqnarray}\label{B23}
&& \int d^3 \vec{x} {\cal I}_3' e^{2iE_{\text{di}}\vec{ v}\cdot \vec{ x}     } 
=2\pi \int \rho d\rho dz' \frac{{\cal J}_{11}}{r_-r_+} \bigg\{  \Big[
 \left(
\frac{r_\Delta^2}{4}  -z^2 
\right)- \frac{1}{2}\rho ^2 \sin^2 \theta \Big]
\\
&& \qquad\qquad\qquad\times J_0(\delta_\rho) \cos(\delta_z)  +\frac{1}{2} \rho^2 \sin^2 \theta J_2(\delta_\rho)  \cos (\delta_z) + \rho z \sin (2\theta ) J_1(\delta_\rho ) \sin(\delta_z)
\bigg\}\,, \nonumber
\end{eqnarray}
where we have utilized that ${\cal J}_{nm}$ is independent of $\phi$. 
Finally, taking all into account, we have 
\begin{eqnarray}
&&\langle p (\vec{v}\,), \downarrow | \overline{ u} \gamma^1 \gamma_5  d (0)  |n  (- \vec{v}\,), \uparrow  \rangle
\nonumber\\
&&\qquad= \frac{5}{3}{\cal N}_n^2\int d^3 \vec{x}_\Delta\int  d^3\vec{x} {\cal I} e^{2iE_{\text{di}} \vec{ v}\cdot \vec{ x}     }\big(
D_u^v (\vec{x}_\Delta)
\big)^2\nonumber\\
&&\qquad = \gamma 
\frac{5}{3}{\cal N}_n^2\int d^3 \vec{x}_\Delta\int  d^3\vec{x}\left(  {\cal I}_1 +{\cal I}_2 +{\cal I}_3' \right)  e^{2iE_{\text{di}} \vec{ v}\cdot \vec{ x}     }\big(
D_u^v (\vec{x}_\Delta)
=1.31 \overline{u}_p u _n,
\end{eqnarray}
where the last equation is evaluated numerically by collecting Eqs.~\eqref{B16} and \eqref{B23}, and taking $v\to 0$. 
Comparing it to  Eq.~\eqref{formfactorss}, we find that $F_1^A=1.31$, which is the desired result.

On the other hand,  the tensor form factors can be obtained directly by  the substitutions
\begin{equation}
	(d,u)\to (b,s)\,,~~~
{\cal G} \to \left(
\begin{array}{cc}
-q_0\hat{x}' \cdot \vec{\sigma} &i q_3 \hat{y}' \cdot \vec{\sigma} \\
-q_3i\hat{y}' \cdot \vec{\sigma} &q_0 \hat{x}' \cdot \vec{\sigma} 
\end{array}
\right)\,,~~~~{\cal I} \to -q_0{\cal I}_1 + q_3 {\cal I}_2 '+ q_0{\cal I}_3' \,
\end{equation}
with
\begin{equation}
{\cal I}_2 ' =- i\frac{1}{2} \left[ (j_{0s}^+ j_{1b}^- - j_{1s}^-j_{0b}^+ )\hat{v}\cdot \hat{x}^- - (j_{1s}^+j_{0b}^- - j_{0s}^-j_{1b}^+)\hat{v}\cdot \hat{x}^+\right]
\end{equation}
in Eq.~\eqref{B13}. Note that ${\cal N}_{\text{flip}} = 1 $  for  $\Lambda_b \to \Lambda$.

\end{document}